\documentclass[11pt]{article}
\usepackage{jheppub}
\pdfoutput=1

\usepackage{slashed}
\usepackage{array,multirow}
\usepackage{graphicx,graphics}
\usepackage{amsfonts,amsmath,amssymb, bbold,bbm}
\usepackage{color,xcolor}
\usepackage{tikz} 
\usepackage{enumitem}
\usepackage{cleveref}

\newcommand{\be}{\begin{equation}}
\newcommand{\ee}{\end{equation}}
\renewcommand\({\left(}
\renewcommand\){\right)}
\renewcommand\[{\left[}
\renewcommand\]{\right]}
\newcommand{\dd}{{\rm d}}

\newcommand{\e}{{\rm e}}

\newcommand\eps{\epsilon}

\def\E{\mathcal{E}}
\def\L{\mathcal{L}}

\def\D{\mathcal{D}}

\renewcommand{\Re}{{\rm Re}}
\renewcommand{\Im}{{\rm Im}}

\def\nn{\nonumber}

\usepackage{tikz}
\usetikzlibrary{patterns}
\usetikzlibrary{arrows,shapes}
\usetikzlibrary{trees}
\usetikzlibrary{matrix,arrows} 				% For commutative diagram
\usetikzlibrary{positioning}				% For "above of=" commands
\usetikzlibrary{calc,through}				% For coordinates
\usetikzlibrary{decorations.pathreplacing}  % For curly braces
\usepackage{pgffor}							% For repeating patterns

\usetikzlibrary{decorations.pathmorphing}	% For Feynman Diagrams
\usetikzlibrary{decorations.markings}
\tikzset{
	vector/.style={decorate, decoration={snake}, draw},
	provector/.style={decorate, decoration={snake,amplitude=2.5pt}, draw},
	antivector/.style={decorate, decoration={snake,amplitude=-2.5pt}, draw},
	fermion/.style={draw=black, postaction={decorate},
		decoration={markings,mark=at position .55 with {\arrow[draw=black]{>}}}},
	fermionbar/.style={draw=black, postaction={decorate},
		decoration={markings,mark=at position .55 with {\arrow[draw=black]{<}}}},
	fermionnoarrow/.style={draw=black},
	gluon/.style={decorate, draw=black,
		decoration={coil,amplitude=4pt, segment length=5pt}},
	scalar/.style={dashed,draw=black, postaction={decorate},
		decoration={markings,mark=at position .55 with {\arrow[draw=black]{>}}}},
	scalarbar/.style={dashed,draw=black, postaction={decorate},
		decoration={markings,mark=at position .55 with {\arrow[draw=black]{<}}}},
	scalarnoarrow/.style={dashed,draw=black},
	electron/.style={draw=black, postaction={decorate},
		decoration={markings,mark=at position .55 with {\arrow[draw=black]{>}}}},
	bigvector/.style={decorate, decoration={snake,amplitude=4pt}, draw},
}

\preprint{ Nikhef-2021-016}

\title{A different perspective on the vev insertion approximation for electroweak baryogenesis
}

\abstract{In the vev insertion approximation (VIA) the spacetime
  dependent part of the mass matrix is treated as a perturbation. We
  calculate the source terms for baryogenesis expanding both the
  self-energy and propagator to first order in mass insertions, which
  gives the same results as the usual approach of calculating the
  self-energy at second order and using zeroth order propagators. This
  procedure shows explicitly the equivalence between including the
  mass in the free or in the interaction Lagrangian. The VIA source
  then originates from the same term in the kinetic equation as the
  semi-classical source, but at leading order in the derivative
  expansion (the expansion in diamond operators). On top, another type
  of derivative expansion is done, which we estimate to be valid for a
  bubble width larger than the inverse thermal width. This cuts off
  the divergence in the VIA source in the limit that the thermal width
  vanishes.  }

\author[a]{Marieke Postma}
\emailAdd{mpostma@nikhef.nl}
\affiliation[a]{Nikhef, Theory Group, Science Park 105, 1098 XG, Amsterdam, The Netherlands}

\begin{document}

\maketitle

%%%%%%%%%%%%%%%%%%%%%%%%%%%%%%%%%%%%%%%%%%%%%%%%%%
%%%%%%%%%%%%%%%%%%%%%%%%%%%%%%%%%%%%%%%%%%%%%%%%%%
%%%%%%%%%%%%%%%%%%%%%%%%%%%%%%%%%%%%%%%%%%%%%%%%%%

\section{Introduction}

In electroweak baryogenesis the matter-antimatter asymmetry of the
universe is created during the electroweak phase transition \cite {Bochkarev:1990fx,Cohen:1990py,Cohen:1990it,Turok:1990zg}.  The scenario relies on new physics at the electroweak scale, which can be probed by the LHC \cite{Arhrib:2013oia, Chen:2013rba, Chang:2017ynj, CMS:2016tgd,Englert:2014uua, Brivio:2017vri,Han:2009ra, Boudjema:2015nda, Ellis:2015dha, Askew:2015mda, Demartin:2015uha},  electric dipole moment measurements \cite{ACME:2018yjb,Chupp:2017rkp,Balazs:2016yvi,Chien:2015xha,Cirigliano:2016nyn, deVries:2017ncy,Andreev:2018ayy}, and possibly future gravitational wave observatories \cite {caprini2016science,Caprini:2019egz,Crowder:2005nr}. This requires accurate theoretical predictions.

The dynamics of the phase transition is a non-equilibrium,
non-perturbative, and finite-temperature process, and to arrive at
transport equations describing the (phase space) densities of the
particles in the plasma various approximations and expansions have to
be made
\cite{Garbrecht:2018mrp,Joyce:1994zn,Joyce:1994zt,Cline:2000nw,Kainulainen:2001cn,Kainulainen:2002th,Prokopec:2003pj,Konstandin:2013caa,Konstandin:2004gy,Prokopec:2004ic,Konstandin_2005,Fromme:2006wx,Cline:2020jre,Cirigliano:2011di,Cirigliano:2009yt,Riotto:1995hh,Riotto:1997vy,Lee:2004we,Carena:2000id}. Different
theoretical approaches have been shown to give predictions for the
asymmetry that may vary by more than an order of magnitude
\cite{Konstandin:2013caa,Cline:2020jre}.

In this paper we take a closer look at the vev insertion approximation
(VIA), in which the source term in the transport equations is derived
expanding in the spacetime dependent part of the mass matrix
\cite{Riotto:1995hh,Riotto:1997vy,Lee:2004we}.  It is, however, not
clear how exactly the sources in VIA are related to the source terms
derived in other approximation schemes. In the standard derivation the
VIA source derives from the collision term in the kinetic equations,
which is puzzling, as this contribution is absent in approaches that
do not expand in mass perturbations.  Moreover, it has been observed
that the VIA source term diverges (in the degenerate mass limit) in
the limit that the thermal width is taken to zero, while it is not
clear what the physical origin of this enhancement is
\cite{Cline:2020jre}.

The VIA source can be derived from the Schwinger-Dyson equations,
which in turn can be reformulated in terms of the Kadanoff-Baym (KB)
equations. The spacetime dependent mass is treated as an interaction,
resulting in non-zero self energies. In the standard derivation the
self-energy is calculated to second order in VIA, that is, from
diagrams with two insertions of the mass, while the propagators are
those of the free theory (dressed with thermal corrections)
\cite{Riotto:1995hh,Riotto:1997vy,Lee:2004we}. Using these expressions
in the collision term in the kinetic equation then gives the source
term at second order in VIA.  However, the self-energy diagram is not
one-particle irreducible (1PI) in that it can be split by cutting a
single line. In the usual resummation of self-energies in the
Schwinger-Dyson equations one would only include the 1PI diagrams, to
avoid double counting.  At second order in VIA we still expect to get
the correct results, but this approach may obscure the origin of the
source term.

Instead, we will calculate both the self-energy and the propagator at
first order in VIA to calculate the source term at 2nd order, and show
that this gives the same final result for the source term as the
standard approach.  For simplicity we work with a scalar toy model,
but we expect the results to  be straightforwardly generalizable to
fermions as well. At first order the (hermitian) self-energy is
nothing but a mass insertion. This immediately makes it clear that it
is equivalent to (1) include the mass term $\delta m^2(z)$ in the free
Lagrangian, and then expand the KB equations in $\delta m^2(z)$; or
(2) include the mass term in the interaction Lagrangian,  and then
expand the KB equations in first order self-energy $\Pi^h_{(1)}$. The
VIA source thus derives from the same term $\propto [-i\delta m^2+\Pi^h,
G^\lambda]$, with $G^\lambda$ the Wightman function, in the kinetic KB equation as the classical source term.

The VIA source is obtained at leading order in the derivative
expansion, that is, at leading order in an expansion of the diamond
operator of the Wigner space KB equation
\cite{Riotto:1995hh,Riotto:1997vy,Lee:2004we}. To evaluate the
resulting integral expression, on top another type of derivative
expansion is done, which assumes the mass insertion -- and thus the
bubble wall background -- varies slowly compared to the typical scales
that dominate the integral. We estimate this expansion to be valid for
bubble wall widths $L_w \gg \max \( \Gamma_T^{-1},\bar m_T^{-1}\)$, with
$\Gamma_T$ thermal width and $\bar m_T^2$ the constant diagonal mass term
including thermal corrections. This cuts off the earlier mentioned
divergence in the source term for $\Gamma_T \to 0$, as in this limit the approximations used are no longer valid.

This paper is organized as follows. In the next \cref{s:model} we
briefly introduce our scalar toy model. All results are formulated in
general terms, and can easily be adapted to other set-ups. \Cref{s:KB}
reviews the derivation of the KB equations from the Schwinger-Dyson
equations, and the transformation to Wigner space. In \cref{s:via} we
then focus on the derivation of the source in
VIA. \Cref{s:self-energy} discusses the the self-energy at first order
in VIA and the equivalence between absorbing the mass perturbation in
the free or in the interaction Lagrangian. We calculate the source
from the self-energy and Green's functions both expanded to first
order in \cref{s:first}, and show that the results are equivalent to
the usual approach in \cref{s:second}. Finally, we estimate the
validity of the derivative-like expansion done to arrive at the final
VIA results in \cref{s:valid}.  We end with concluding remarks.

%%%%%%%%%%%%%%%%%%%%%%%%%%%%%%%%%%%%%%%%%%%%%%%%%%
%%%%%%%%%%%%%%%%%%%%%%%%%%%%%%%%%%%%%%%%%%%%%%%%%%
%%%%%%%%%%%%%%%%%%%%%%%%%%%%%%%%%%%%%%%%%%%%%%%%%%
\section{Scalar toy model}
\label{s:model}

For simplicity we will consider a scalar model in this paper, which
avoids complication with spin projections,
but we expect the results
can be rather straightforwardly generalized to fermionic models as
well.

Consider then a two-flavor scalar model with CP violating couplings to the
bubble background \cite{Cirigliano:2011di,Cirigliano:2009yt}. The quadratic Lagrangian for the flavor doublet $\phi =( \phi_L \; \phi_R)^T$ is
\be
\L= (\partial_\mu \phi)^\dagger (\partial^\mu \phi) -\phi^\dagger M^2(z)  \phi 
\label{L}
\ee
with mass matrix
\be
M^2 =M_D^2 + \delta M^2(z) =
\(
\begin{array}{cc}
  m_L^2 & 0\\
 0 & m_R^2
\end{array}
\)+
\(
\begin{array}{cc}
  0 & \delta m^2(z)\\
 ( \delta m^2(z))^* & 0
\end{array}
\).
\label{mass}
\ee
The flavor-diagonal masses $m^2_{L,R}$ are constant, while the
off-diagonal mass $\delta m^2 =|\delta m^2| \e^{i\theta}$ depends on
the bounce solution $v_b =v_b(z)$ describing the bubble wall. In VIA
the off-diagonal term is treated as an interaction, and the mass and
flavor eigenstates coincide for the free Lagrangian.  In addition,
$\phi$ interacts with the degrees of freedom in the plasma. These
plasma effects are incorporated dressing the Green's functions with a
non-zero self-energy. Essential for the VIA mechanism is that this
self-energy correction is complex, and gives apart from a thermal mass
also a thermal width correction \cite{Riotto:1995hh,Riotto:1997vy}.

% An explicit realisation of this type of set-up can be found e.g. in
% SUSY extensions of the SM such as the NMSSM; then $m_{L,R}^2$ are the soft SUSY breaking masses and $\delta m^2$ can arise from a coupling to the singlet-Higgs sector.

For future reference we define the Klein-Gordon operator ${\cal D}(u)$ via
\begin{align}
S = -\int \dd^4z \,
\phi(z)^\dagger \D(z) \phi(z)
       =-\int \dd^4z\,
  \phi(z)^\dagger ( \partial_z^2 +M^2(z)) \phi (z)  .
  \label{D}
\end{align}
%

%%%%%%%%%%%%%%%%%%%%%%%%%%%%%%%%%%%%%%%%%%%%%%%%%%
%%%%%%%%%%%%%%%%%%%%%%%%%%%%%%%%%%%%%%%%%%%%%%%%%%
%%%%%%%%%%%%%%%%%%%%%%%%%%%%%%%%%%%%%%%%%%%%%%%%%%

\section{Kadanoff-Baym equations}
\label{s:KB}

In this section we briefly review the derivation of the Kadanoff-Baym
(KB) equations that govern the quantum dynamics of the system. This
also serves to set the notation.

%%%%%%%%%%%%%%%%%%%%%%%%%%%%%%%%%%%%%%%%%%%%%%%%%%
\subsection{CTP formalism and Green functions}

The Closed Time Path (CTP) or Schwinger-Keldysh formalism can describe
finite-temperature and non-equilibrium quantum systems \cite{Schwinger:1960qe,Keldysh:1964ud,Chou:1984es}. The path integral is defined along a contour ${\cal C}$ that starts from an initial time $t_0 \to -\infty$ to time $t = \infty$ and then back, which can be split in a ${\cal C}_+$ and ${\cal C}_-$ branch. The scalar Green function defined via
\be
  G(u,v)
 = \langle \Omega| T_{\cal C} \, \phi(u) \phi^\dagger(v) |\Omega\rangle,
\ee
with $T_{\cal C}$ denoting time-ordering  along the contour ${\cal C}$,
can be split depending on which branch the time-arguments of the fields lie. We use the notation that $G^t = G^{++}, \, G^{\bar t} =G^{--},\, G^> = G^{-+}, \, G^< = G^{+-}$.

The retarded and advanced propagators, and hermitian and anti-hermitian propagators are
\begin{align}
  G^r &\equiv G^t - G^< = G^> - G^{\bar t} ,&G^a &\equiv G^t - G^>  =  G^< - G^{\bar t},
\nonumber\\
    G_h &\equiv \frac 12 (G^r+G^a)= \frac 12 (G^t-G^{\bar t}),& G_{ah} &\equiv \frac {1}{2i} (G^a-G^r)
    =        \frac {i}{2} (G^>-G^<).
\label{Gdefs}
\end{align}
%
%Further relation between the Green functions
% %
% \begin{eqnarray}
%   G^t(u,v)       &=& \theta(u_0-v_0) G^>(u,v) + \theta(v_0-u_0)G^<(u,v)
% \nonumber\\
% %
%   G^{\bar{t}}(u,v) &=& \theta(u_0-v_0) G^<(u,v) + \theta(v_0-u_0)G^>(u,v)
% \,,
% \end{eqnarray}
% %
% and
%
A further useful relation is
\be
G^t (u,v)+G^{\bar t} (u,v)= G^>(u,v) + G^<(u,v).
\label{sumG}
\ee
The hermiticity properties are $G^\lambda(u,v)^\dagger = G^\lambda(v,u)$ with $\lambda = >,\,<$ and $G^t(u,v)^\dagger = G^{\bar t}(v,u)$.

Similarly, one can define the advanced/retarded and hermitian/anti-hermitian scalar self-energies in terms of $\Pi^{ab}$, with $a,b =+,\,-$.

%%%%%%%%%%%%%%%%%%%%%%%%%%%%%%%%%%%%%%%%%%%%%%%%%%%%%%%%%%%

\subsection{From Schwinger-Dyson to Kadanoff-Baym}

The starting point are the Schwinger-Dyson equations, derived form the 2PI effective action in \cite{Prokopec:2003pj,Garbrecht:2018mrp}, which can be written in the form
\begin{align}
     \D (u) G^{ab}(u,v)
  &= -ai\delta_{ab}\delta^4(u-v) 
   -i    \sum_c c \int d^4w \, \Pi^{ac}(u,w)G^{cb}(w,v) ,\nn \\
  G^{ab}(u,v)  \stackrel{\leftarrow}{\D} (v)&= -i a\delta_{ab}\delta^4(u-v)  -i    \sum_c c \int d^4w \, G^{ac}(u,w) \Pi^{cb}(w,v),
\label{eq_Gab}                                     
\end{align}
with ${\cal D}$ the Klein-Gordon operator \cref{D} and $a,b =+,\,-$.
For $\lambda = >,<$ they become
\begin{align}
 \partial_u^2 G_{uv}^\lambda&=- M^2_u
                              G_{uv}^\lambda
                                                      -i  \int \dd^4 w\,
\( \Pi^h_{uw}  G^\lambda_{wv} +\Pi^\lambda_{uw} \ G^h_{wv} +\frac12\(
                                       \Pi^>_{xz}  G^<_{zy} -\Pi^<_{xz}  G^>_{zy} \) \),
                                       \nn \\
  \partial_v^2 G_{uv}^\lambda&=-   G_{uv}^\lambda M^2_v                          
-i  \int \dd^4 w\,
\( G^h_{uw}  \Pi^\lambda_{wv} +G^\lambda_{uw} \Pi^h_{wv} +\frac12\(
                                       G^>_{xz}  \Pi^<_{zy} -G^<_{xz}  \Pi^>_{zy} \) \),
\label{KB}
\end{align}
where we introduced the shorthand notation $A_{uw} = A(u,w)$.
Now add and subtract these equations to obtain the anti-hermitian and
hermitian part, which correspond to the constraint and kinetic
equation respectively:
\begin{align}
   (\partial^2_u  +\partial^2_v ) G^\lambda_{uv}   & =
   - \{ M^2, G^\lambda\}_{uv}  -i  \int \dd^4 w \,\( \{\Pi^h , G^\lambda\}
  +\{\Pi^\lambda, G^h\}  +{\cal C}^-_{\rm coll} 
                                                      \)_{uv}                                                
  \label{eq_plusmin1} \\
    (\partial^2_u  -\partial^2_v ) G^\lambda_{uv}   & =
  -  [M^2, G^\lambda]_{uv}-i  \int \dd^4 w \,\(  [\Pi^h , G^\lambda]
  +[\Pi^\lambda, G^h]  +{\cal C}^+_{\rm coll} 
                                                      \)_{uv}
   \label{eq_plusmin2}
\end{align}
with the shorthand notation $\[\Pi,G\]_{uv} = \Pi_{uw} G_{wu}-G_{uw}\Pi_{wu}$ and similar for the anti-commutators. The collision term is
\be
({\cal C}^\pm_{\rm coll})_{uv}= \frac12\([\Pi^> ,  G^< ]^\pm -[\Pi^< ,  G^>  ]^\pm\) _{uv}
\label{coll}
\ee
with $[..,..]^+ =\{..,..\}$ anti-commutator and $[..,..]^- =[..,..]$ commutator.  Here we used
\cref{Gdefs} to rewrite the right hand side
\begin{align}
  \Pi^> G^t -\Pi^{\bar t} G^>
 %  &= \Pi^> G^a +\Pi^r G^> \nn\\
% & =\Pi_h G^> + \Pi^> G_h +\frac12 \( (\Pi^r-\Pi^a) G^>+ \Pi^>(G^a-G^r)\) \nn \\
                              &= \Pi_h G^> + \Pi^> G_h +\frac12\( \Pi^> G^< -\Pi^< G^>\) ,
\nn \\
  \Pi^t G^< -\Pi^{<} G^{\bar t} % &= \Pi^r G^< +\Pi^< G^a
                                %    \nn \\
                              &= \Pi_h G^< + \Pi^< G_h +\frac12\( \Pi^> G^< -\Pi^< G^>\) .
\end{align}
\Cref{eq_plusmin1,eq_plusmin2} are the position space Kadanoff-Baym (KB) equations.

%%%%%%%%%%%%%%%%%%%%%%%%%%%%%%%%%%%%%%%%%%%%%%%%%%%%%%%%
\subsection{Wigner representation}

To do a gradient expansion in the slowly varying bubble background, it is useful to define the relative and center-of-mass coordinates
\be
x= \frac12(u+v),\quad r = u-v  \qquad \Leftrightarrow  \qquad
u = x+\frac12 r, \quad v =  x-\frac12 r.
\label{CM_coord}
\ee
%
%such that $\partial_u=\frac12 \partial_x +\partial_r$,  $ \partial^2_u
% +\partial^2_v= \frac12 \partial_x^2 +2 \partial_r^2$ and $\partial^2_u
% - \partial^2_v = 2 \partial_x \cdot \partial_r$.
We use the notation for a general function $A(u,v)$
\be
A(u,v) =  A(x+\frac12 r, x-\frac12 r)  \equiv \bar A(r,x).
\ee
The Wigner transform is defined as the Fourier transform with respect to the relative coordinate
\be
A(k,x) = \int \dd^4 (u-v) \, \e^{ik. (u-v)} A(u, v) = \int \dd^4 r \,
\e^{i k.r} \bar A(r,x).
\ee
Now integrate \cref{KB} with a factor $\int \dd^4(u-v) \e^{i k.(u-v)}$, and add and subtract the two equations to extract the hermitian and anti-hermitian parts
\begin{align}
  \(\frac12 \partial_x^2 -2 k^2\) G^\lambda(k,x)& = - i\e^{-i\diamond}\(
\{-i M^2+ \Pi^h , G^\lambda \}+ \{\Pi^\lambda , G^h\} +{\cal C}^-_{\rm coll} \),
    \label{constraint}
  \\
 2i k\cdot \partial_x G^\lambda(k,x)  & = i\e^{-i\diamond}\(
[-i M^2+  \Pi^h , G^\lambda ]+ [\Pi^\lambda ,\ G^h] +{\cal C}^+_{\rm coll} \).
    \label{kinetic}
\end{align}
where all $G^i,\, \Pi^i$ are a function of $(k,x)$, and $M^2$ a function of $(x)$.  The collision term \cref{coll} in Wigner space is
\be
{\cal C}^\pm_{\rm coll} (k,x)= \frac12\([\Pi^> (k,x),  G^< (k,x)]^\pm -[\Pi^< (k,x),  G^> (k,x) ]^\pm\).
\label{collW}
\ee
The diamond operator is defined as
\be
\diamond \big(A(k,x) B(k,x) \big)= \frac12 \big(\partial_x A(k,x) \cdot\partial_k B(k,x) -\partial_k A(k,x) \cdot \partial_x B(k,x)\big).
\ee
Some useful relations involving the diamond operator, used in the above derivation, are collected in \cref{A:diamond}.

\Cref{constraint,kinetic} are the Wigner space Kadanoff-Baym equations, corresponding to the constraint and kinetic equation respectively.

%%%%%%%%%%%%%%%%%%%%%%%%%%%%%%%%%%%%%%%%%%%%%%%%
%%%%%%%%%%%%%%%%%%%%%%%%%%%%%%%%%%%%%%%%%%%%%%%%
%%%%%%%%%%%%%%%%%%%%%%%%%%%%%%%%%%%%%%%%%%%%%%%%

\section{VEV insertion approximation}
\label{s:via}

In the vev insertion approximation (VIA) the off diagonal and
space-time dependent parts of the mass matrix, $\delta M^2$ in
\cref{mass}, are treated as a perturbation.  Flavor oscillations and
coherence effects are neglected. The source terms are calculated
expanding the KB equations in the number of mass insertions, which
should converge for sufficiently small enough mass perturbations
$\delta m^2 /(\bar m_T \Gamma_T)\ll 1$, with
$\bar m_{T,i}^2$ the flavor diagonal mass including thermal
corrections and $\Gamma_T$ the thermal width \cite{Postma:2019scv}.

In the usual derivation of the source term in VIA the self-energy is
calculated for two mass insertions, while the Green's functions of the
free theory (dressed with thermal corrections) are used
\cite{Riotto:1995hh,Riotto:1997vy,Lee:2004we}. As we will show in
section \cref{s:second} this gives the same result as calculating the
both the self-energy and the Green's function to first order in VIA.
We first follow the latter approach.

%%%%%%%%%%%%%%%%%%%%%%%%%%%%%%%%%%%%%%%%%%%%%%%%%%%%%%%%%%
\subsection{Self-energy at first order in VIA}
\label{s:self-energy}

It is clear that we can either (1) include $\delta M^2$ in the free Lagrangian, derive the KB equations, and expand in $\delta M$, or (2) include  $\delta M^2$ in the interaction Lagrangian which result in a non-zero self-energy, derive the KB equations and expand in powers of the self-energy.  These are two different descriptions of the same expansion of the same physical system.  We proceed with the latter approach and  set $\delta M^2=0$ in the KB equations, and incorporate the mass insertions via the self-energy. 

At first order in VIA (one mass insertion) the non-zero self energies in position space are
\be
\Pi_{ij,(1)}^{ab}(u,v)  = f_{ij}^{ab}(u) \delta^4(u-v),
\ee
with
\begin{align}
  f_{LR}^{++}(u)  &= -i \delta m^2(u),&
  f_{LR}^{--}(u,)  &= i \delta m^2 (u),\nn\\
  f_{RL}^{++}(u) &=-i (\delta m^2(u))^* ,&
 f_{RL}^{--}(u,v) &=i (\delta m^2 (u) )^*.
\label{Pi_0}
\end{align}
The vertices on the ${\cal C}^-$ branch pick up an additional minus
sign.  Throughout we will use $a,b,.. =+,-$ for CTP contour indices
and $i,j,.. = L,R$ for flavor indices. The subscript $(\#)$ on
$\Pi^{ab}_{ij}$ and $G^{ab}_{ij}$ refers to the number $\#$ of mass
insertions.  Flavor diagonal self-energies vanish at this order, as
well as $\Pi^\lambda$, as vertices always are inserted on the same
branch of the CTP contour. We note that the hermitian self-energy
\be
\Pi_{LR,(1)}^{h}(u,v) =\frac12\(\Pi_{LR,(1)}^{++}(u,v) -\Pi_{LR,(1)}^{--}(u,v)\)= -i \delta m^2(v) \, \delta^4(u-v)
\label{Pi_1}
\ee
is proportional to the mass perturbation, whereas the orthogonal
combination $\propto ( \Pi_{LR,(1)}^{++} +\Pi_{LR,(1)}^{--})=0$
vanishes. If we insert these results in the position space KB
\cref{eq_plusmin1,eq_plusmin2}, the integral over $\Pi_h$ can be done
with the delta-function, and using \cref{Pi_0}, the result is
\begin{align}
   (\partial^2_u  +\partial^2_v ) G^\lambda_{uv}   & =
   - \{M_D^2+\delta M^2, G^\lambda\}_{uv},  \nn \\
    (\partial^2_u  -\partial^2_v ) G^\lambda_{uv}   & =
    [M_D^2+\delta M^2, G^\lambda]_{uv}.
\end{align}
This indeed gives exactly the same equations as one would have obtained by absorbing $\delta M^2$ in the free Lagrangian and setting $\Pi_h=0$.

From the KB equations in Wigner space \cref{constraint,kinetic} it is even more transparant that both procedures give the same result, as these only depend on the combination $(-i \delta M^2+ \Pi^h)$.  The Wigner space self-energy is
\be
\Pi_{ij,(1)}^{ab} (k,x)=\int \dd^4 r \,\e^{ikr} f^{ab}_{ij} (x+\frac12 r)\delta(r) =f^{ab}_{ij} (x) ,
%\qquad \Pi_{LR,(1)}^{h} =-i \delta m^2(x)
\label{PI_0_W}
\ee
and thus $\Pi_{LR,(1)}^{h} =-i \delta m^2(x)$,
which depends on the collective coordinate $x = u= v$, but not on the momenta conjugate to the relative coordinates.

The above implies that the source term in VIA can be equally derived from an expansion in $(-i\delta M^2)$ as from an expansion in $\Pi_h$.  The source term thus should arise from
\be
2i k\cdot \partial_x G^\lambda(k,x)   = i\e^{-i\diamond}
[-i(M_D^2(x)+\delta M^2(x))+  \Pi^h(x) , G ^\lambda  (k,x)]
\label{WKB}
\ee
in the kinetic equation. This is the same term that gives rise to the
semi-classical force \cite{Joyce:1994zt,Joyce:1994fu,Huber:2000ih,Cline:1997vk,Cline:2000nw}.  The KB equations resums all mass insertions; instead in VIA the kinetic equation is solved perturbatively expanding in mass insertions.

%%%%%%%%%%%%%%%%%%%%%%%%%%%%%%%%%%%%%%%%%%%%%%%%%%%%%%%%%%%%%%%
\subsection{Source term from first order self-energy}
\label{s:first}

To derive the transport equations at 2nd order in the VIA expansion,
we perform a derivative expansion of the kinetic equations in Wigner
space, and then transform the results back to position space. The
propagators are most easily derived directly in position space from
the Schwinger-Dyson equations.  We will calculate the source at 2nd
order expanding both the self-energy and Green's functions to first
order in VIA.  The self-energy was discussed in the previous section.

Setting $\delta M^2 =0$ in the free Lagrangian, the mass matrix is flavor diagonal, and the zeroth order in VIA constraint equation for the Green's function becomes
\begin{align}
 (\partial_u^2 +m_i^2)  G^{ab}_{ij,(0)}(u-v) = -i a \delta_{ab} \delta^4(u-v)
\end{align}
for $i=L,R$ flavors. The mass matrix is diagonal and the solutions are
the usual free thermal propagators \cite{Landsman:1986uw}. Coherence
effects are neglected ($f_{LR}$ distributions are set to zero) and the
off-diagonal propagators vanish. In VIA the propagators are dressed
with a non-zero self-energy to account for the interaction with the
thermal plasma \cite{Parwani:1991gq}, which amounts to adopting the
spectral function \cite{Riotto:1995hh}
\be
\rho(k) = i \[ \frac{1}{(k^0 +i\eps +i \Gamma_T)^2 - \omega^2}
  -\frac{1}{(k^0 -i\eps -i \Gamma_T)^2 - \omega^2}\],
\label{spectral}
\ee
with $\omega^2 = \vec k^2 +m_i^2+m_{T,i}^2$ and $m^2_T,\,\Gamma_T $ the
thermal mass and width respectively \cite{Riotto:1995hh}. For
vanishing plasma corrections $m^2_T,\,\Gamma_T \to 0$ this reduces to
the free spectral function. 

At first order in VIA the Schwinger-Dyson equation reads
\begin{align}
  G^{ab}_{(1),ij}(u,v) &= \sum_{cd} c d \int \dd^4 z_1 \int \dd^4 z_2\, G^{ac}_{(0),ik}(u-z_1) \Pi^{cd}_{(1),kl}(z_1-z_2) G^{db}_{(0),lj}(z_2-v)  .                     
\end{align}
Using the non-zero first order self-energy  \cref{Pi_0} and zeroth order Green's function this gives
\begin{align}
  G_{(1),ij}^{++}(u,v) &= \sum_{cd}  \int \dd^4 z_1 \int \dd^4 z_2\, G^{+-}_{(0),ii}(u-z_1) \Pi^{--}_{(1),ij}(z_1-z_2) G^{-+}_{(0),jj}(z_2-v)
  \nn \\&=\int \dd^4 z \,G^{+-}_{(0),LL}(u-z) f^{--}_{ij}(z) G^{-+}_{(0),RR}(z-v) , \nn \\
  G_{(1),ij}^{--}(u,v) &=\int \dd^4 z \,G^{-+}_{(0),ii}(u-z) f^{++}_{ij}(z) G^{+-}_{(0),jj}(z-v),
                         \label{Gt_first}
\end{align}
for $i\neq j$, and as before $i,j=L,R$. 

The source term can be derived from the kinetic equation \cref{kinetic}, which at 2nd order becomes
\begin{align}
  k\cdot \partial_x  \(G^>+G^<\)_{ij}   &= \frac12 \e^{-i\diamond}
                                                    \( [\Pi^h  , G^t+G^{\bar t} ]+
                                          [\Pi_i^t+\Pi_i^{\bar t} ,\ G^h ] +{\cal C}^+_{\rm coll} \)_{ij} \nn\\
  & \stackrel{(2)}{=}  \frac12 \e^{-i\diamond}
                                                    \(  [\Pi_{(1)}^h  , G_{(1)}^t+G_{(1)}^{\bar t}]\) \delta_{ij}.
  \label{kineticplus}
\end{align}
On the first line we used \cref{sumG}, and on the 2nd line we gave the 2nd order VIA contribution originating from multiplying a 1st order 1PI self-energy with a 1st order propagator.  Integrating over momenta gives the derivative of the number current (particles minus \nobreak{antiparticles}) density for the left hand side. Using \cref{PI_0_W} to write $\Pi_{(1)}^h(k,x)  = f^h(x)$, which only depends on the collective coordinate $x$, this gives
 \begin{align}
\partial_\mu j_{(2),i}^\mu(x)&\equiv
 \int \frac{\dd^4 k}{(2 \pi)^4}  \, k^\mu \, 
            \left(  G_{(2)}^< (k,x)    +  G_{(2)}^> (k,x) \right)_{ii} \nn\\
 &= \frac12\int \frac{\dd^4 k}{(2 \pi)^4}  \, \e^{-i\diamond}
   \(  [f^h(x)  , G_{(1)}^t(k,x)+G_{(1)}^{\bar t}(k,x)]\)_{ii}.
   \label{current1}
\end{align}
The right hand side is known as the source term $S$, i.e., we write
$\partial_\mu j_{(2),i}^\mu(x) =S_{(2),i}$ at second order in VIA.  The first two terms in the derivative expansion, corresponding to an expansion of $\e^{-i\diamond} = 1 - i \diamond +{\cal O}(\diamond^2)$, are
\begin{align}
 S_{(2),i}(x)|_{\textsc{LO}} &= \frac12\int \frac{\dd^4 k}{(2 \pi)^4}  \, 
     [f^h(x)  , G_{(1)}^t(k,x)+G_{(1)}^{\bar t}(k,x)]_{ii} ,\nn \\
S_{(2),i}(x)|_{\textsc{NLO}} &=-\frac{i}{2}
\{\partial_x f^h(x)  , \partial_k(G_{(1)}^t(k,x)+G_{(1)}^{\bar t}(k,x))\}_{ii},
\end{align}                                  
with LO and NLO denoting the leading and next-to-leading order in the derivative expansion.
To transform back the results to coordinate space we use that
\begin{align}
\int \frac{\dd^4 k}{(2 \pi)^4}  \, f^h(x) G(k,x) &= \int \frac{\dd^4 k}{(2 \pi)^4}  \int \dd^4 r \, \e^{ikr} f^h(x)\bar G(r,x)
                             =f^h(x) \bar G(0,x) = f^h(u) G(u,u), \nn \\
  \int \frac{\dd^4 k}{(2 \pi)^4}  \,\partial_x  f^h(x) \partial_k G(k,x)
                                                 &= \int \frac{\dd^4 k}{(2 \pi)^4}  \int \dd^4 r \, (ir)\partial_x f^h(x)\bar G(r,x) \e^{ikr} =0.
\label{der_vanish}                                                                               
\end{align}
Thus only the leading order term gives a non-zero result; all higher order terms in the derivative expansion vanish as they are proportional to $r^n$ (and the $k$-integral gives a $\delta(r)$). We thus only get a source term at leading order in the derivative expansion
 $S_{(2),ii}(u)= S_{(2),ii} \big|_{\textsc{LO}}$
 with
 \begin{align}
 S_{(2),ii} 
   &= \frac12\[f^h(u)  ,G_{(1)}^t(u,u)+G_{(1)}^{\bar t}(u,u)\]_{ii} \nn \\
   &=  {\rm Re} \int \dd^4 w \Big( f^{\bar t}_{ij}(u)G^<_{(0),jj} (u,w)   f^{ t}_{ji} (w) G^>_{(0),ii} (w,u) \nn\\
   & \hspace{2.1cm}- f^{ t}_{ij}(u)G^>_{(0),jj} (u,w)   f^{\bar t}_{ji} (w) G^<_{(0),ii} (w,u)\Big).
     \label{source0}
 \end{align}
On the 2nd line we rewrote the first order Green's function in terms
of the lowest order Green's functions using \cref{Gt_first}.  This is the same result as in the standard calculation, as shown in the next subsection.

%%%%%%%%%%%%%%%%%%%%%%%%%%%%%%%%%%%%%%%%%%%%%%%%%%%%%%

 \subsection{Source term from 2nd order self-energy}
\label{s:second}

The standard approach the source is derived using the 0th order Green's functions and the 2nd order self-energy.  In this subsection we briefly review this derivation, and show that the resulting source term is the same as derived in the previous subsection.

The kinetic equation \cref{eq_plusmin2} for $G^> + G^< = G^t +G^{\bar t}$ can be written in the form
\begin{align}
    (\partial^2_u \! -\!\partial^2_v ) ( G^t \!+\!G^{\bar t})_{uv}   %&=\nn \\
% &=-i  \!\int \! \dd^4 w \( [\Pi^h , G^t+G^{\bar t}]
%   +[\Pi^t+\Pi^{\bar t}, G^h]+  \{\Pi^> ,  G^< \} -\{\Pi^< ,  G^>  \}
%                                                       \)_{uv}                                                
%   \nn \\
 &=-i  \!\int \!\dd^4 w \( [\Pi^{++} , G^{++}]-[\Pi^{--} , G^{--}]
  +  \{\Pi^{-+} ,  G^{+-} \}-\{\Pi^{+-}, G^{-+}\} 
                                                      \)_{uv}       
   \label{KB_tt}
\end{align}
where for notational convenience we suppressed flavor indices. The last two terms on the right hand side originate from the collision term. At second order this becomes
\begin{align}
 (\partial^2_u  -\partial^2_v ) ( G^t+G^{\bar t})_{uv}   %&=\nn \\
&\stackrel{(2)}{=}
 -i  \int \dd^4 w \,\( [\Pi_{(1)}^{++} , G_{(1)}^{++}]-[\Pi_{(1)}^{--} , G_{(1)}^{--}]
\)_{uv}       \nn \\
&\stackrel{(2)}{=}
 -i  \int \dd^4 w \,\( 
  -\{\Pi_{(2)}^{+-}, G_{(0)}^{-+}\}+  \{\Pi_{(2)}^{-+} ,  G_{(0)}^{+-} \} 
                                                                             \)_{uv},
                       \label{equal}
\end{align}
where the first line is the approach of the previous subsection, and the 2nd line the standard derivation. Including both would be double counting.  To show that both expressions are indeed the same, we start with the self-energy at 2nd order in VIA
\be
\Pi_{(2),ii}^{ab}(u,v) = cd \int \dd^4 z_1 \int \dd^4 z_2 \,\Pi_{(1),ij}^{ac}(u,z_1) G_{(0),jj}^{cd}(z_1,z_2) \Pi_{(1),ji}^{db}(z_2,v).
\label{Pi2}
\ee
\begin{figure}[t]
\begin{center}
\begin{tikzpicture}[line width=1.5 pt, scale=1]
  \draw(-0.4,1) -- (3.1,1);
 \draw (45:0.15)+(0.3,1) -- (0.3,1);
  \draw (-45:0.15)+(0.3,1) -- (0.3,1);
 \draw (135:0.15)+(0.3,1) -- (0.3,1);
 \draw (-135:0.15)+(0.3,1) -- (0.3,1);
\draw (45:0.15)+(2.4,1) -- (2.4,1);
  \draw (-45:0.15)+(2.4,1) -- (2.4,1);
 \draw (135:0.15)+(2.4,1) -- (2.4,1);
 \draw (-135:0.15)+(2.4,1) -- (2.4,1);
\node at (-0.1,1.2) {$+$};
\node at (0.7,1.2) {$+$};	
\node at (2.,1.2) {$-$};
 \node at (2.8,1.2) {$-$};
  \node at (1.3,0.0) {$\underbrace{-\Pi^{++}_{(1),ij} \odot
                            G_{(0),jj}^{+-}\odot\Pi^{--}_{(1),ji}
                          }_{\Pi^{+-}_{(2),ii}}$};
   \node at (3.8,1) {$\odot$};
\draw(4.4,1) -- (6,1);  
 \node at (4.7,1.2) {$-$};	
         \node at (5.7,1.2) {$+$};
         \node at (5.3,0.4)
        {$G^{-+}_{(0),ii} $};
        \node at (6.6,1) {$=$};
        \begin{scope}[shift={(8,0)}]
           \node at (-0.7,1) {$-$};
               \draw(-0.4,1) -- (1,1);
 \draw (45:0.15)+(0.3,1) -- (0.3,1);
  \draw (-45:0.15)+(0.3,1) -- (0.3,1);
 \draw (135:0.15)+(0.3,1) -- (0.3,1);
 \draw (-135:0.15)+(0.3,1) -- (0.3,1);
 \node at (-0.1,1.2) {$+$};
\node at (0.7,1.2) {$+$};	
\node at (.3,0.4)     {$\Pi^{++}_{(1),ij} $};
\node at (1.7,1) {$\odot$};
        \draw(2.3,1) -- (6.1,1);
                 \draw (45:0.15)+(4.1,1) -- (4.1,1);
  \draw (-45:0.15)+(4.1,1) -- (4.1,1);
 \draw (135:0.15)+(4.1,1) -- (4.1,1);
 \draw (-135:0.15)+(4.1,1) -- (4.1,1);
\node at (3.7,1.2) {$-$};
 \node at (4.5,1.2) {$-$};
                \node at (2.6,1.2) {$+$};	
                \node at (5.7,1.2) {$-$};
                %\node at (0.55,1.2) {$+$};	
        \node at (4.1,0.) {$\underbrace{ G_{(0),jj}^{+-}\odot\Pi^{--}_{(1),ji} \odot G_{(0),ii}^{-+} }_{G_{(1),ji}^{++}}$};
     %   \node at (-1.8,1) {$(\Pi_{(2)}^{-+}\times G_{(0)}^{+-})_{ii}\,=$};
        
   %    \node at (5.4,1) {$=$};
\end{scope}
	 \end{tikzpicture}	
\end{center}
\caption{The equivalence  \cref{relation} between $\Pi^{-+}_{(2),ii}\odot G^{+-}_{(0),ii} = -
  \Pi^{--}_{(1),ij}\odot G^{--}_{(1),ji}$,   with $\odot$ denoting
  integration over the internal variable, in terms of Feynman
  diagrams. A mass insertion is represented by a cross, ${\times}$, in the diagrams.}
\label{fig:diagrams}
\end{figure}
It follows that
\begin{align}
  &\int \dd^4 w \, \{\Pi_{(2)}^{+-}(u,w), G_{(0)}^{-+} (w,v) \}\nn \\
  & =- \int \dd^4 w \int \dd^4 z_1 \int \dd^4 z_2 \,
    \Big(\Pi_{(1)}^{++}(u,z_1) G_{(0)}^{+-}(z_1,z_2) \Pi_{(1)}^{--}(z_2,w) G_{(0)}^{-+}(w,v) \nn \\
 &\hspace{4cm}    + G_{(0)}^{-+}(u,w)\Pi_{(1)}^{++}(w,z_1) G_{(0)}^{+-}(z_1,z_2) \Pi_{(1)}^{--}(z_2,v) \Big)
    \nn \\
  & =  -\int \dd^4 z_1 \, \(\Pi_{(1)}^{++}(u,z_1) G_{(1)}^{++}(z_1,v)
    + G_{(1)}^{++}(z_1,v)\Pi_{(1)}^{++}(u,z_1)\),
    \label{relation}
\end{align}
where the expression for the 1st order propagator in \cref{Gt_first} was used. The overall
  minus sign of $\Pi_{(2)}^{+-}$ arises from the definition
  \cref{Pi2}, as the sum over the internal indices \nobreak{$c,d = +,-$} gives
  an extra sign. This expression can be equally understood in terms of
  Feynman diagrams as shown in \Cref{fig:diagrams}, where we also
  reinstated flavor indices.
  Likewise, 
\begin{align}
  \int \dd^4 w \, \{\Pi_{(2)}^{-+}(u,w) ,G_{(0)}^{+-} (w,v)\} = - \!\int \dd^4 z_1 \!\(
  \Pi_{(1)}^{--}(u,z_1) G_{(1)}^{--}(z_1,v)+  G_{(1)}^{--}(z_1,v)\Pi_{(1)}^{--}(u,z_1)\),
\end{align}
which together proof that the two expressions in \cref{equal} above are the same.

Let's briefly review the derivation of the source in terms of the 2nd order self-energy.  The non-zero self-energies are
\begin{align}
\Pi_{(2),ii}^{-+} =  - f_{ij}^{--}(u) G_{(0),jj}^{-+}(u,v) f_{ji}^{++}(v), \quad
\Pi_{(2),ii}^{+-} &=  - f_{ij}^{++}(u) G_{(0),jj}^{+-}(u,v) f_{ji}^{--}(v),
\end{align}
which can be substituted in the kinetic equation \cref{equal}. For the
left hand side we use that\footnote{In
  \cite{Riotto:1995hh,Riotto:1997vy,Lee:2004we} the current
  conservation is derived directly in position space in the limit $v
  \to u$ via $ \lim_{u\to v} i\left(\partial^2_u-\partial^2_v\right)
  \(G^t(u,v)+G^{\bar t}(u,v)\) = 2\partial_\mu j^\mu(x) $.
  % Rewriting the source integral $\int\dd  w^0=2\int\dd  w^0 \, \theta(x^0-w^0)
  % = 2\int_{-\infty}^{t} dw_0$ in  \cref{source2} reproduces their
  % result.
}
\begin{align}
  \frac{i}{2}\left(\partial^2_u-\partial^2_v\right) \(G^t(u,v)+ G^{\bar t}(u,v)\)&
= \partial_{x^\mu} \int \frac{\dd^4 k}{(2\pi)^4}\, k^\mu \e^{-ikr} \( G^<(k,x) +G^>(k,x) \)
                                                                                  % \nn \\ &
                                                                                            =\partial_\mu j^\mu(x)
  \label{cons_curr}                                                                       
\end{align}
with $x =\frac12(u+v)$.
Taking the leading order result in the diamond expansion, corresponding to setting $v =u$ in position space, then gives the source $ \partial_\mu j_{(2)}^\mu(x) |_{\textsc{LO}}=S_{(2),ii}|_{\textsc{LO}}$ with
\begin{align}
S_{(2),ii}|_{\textsc{LO}}
  &= \frac12 \int \dd^4 w \,\(
  -\{\Pi_{(2)}^{+-}, G^{-+}_{(0)}\}+  \{\Pi^{-+}_{(2)} ,  G^{+-}_{(0)} \}
  \)_{uu}
 = \int \dd^4 w  \, \Re
    \( \Pi^{-+}_{(2) } G_{(0)}^{+-}- \Pi_{(2)}^{+-}G^{-+}_{(0)}\)_{uu} \nn \\
  &=  {\rm Re} \int \dd^4 w \Big( f^{\bar t}_{ij}(u)G^<_{(0),jj} (u,w)   f^{ t}_{ji} (w) G^>_{(0),ii} (w,u)
  \nn\\ & \hspace{1.8cm}
            - f^{ t}_{ij}(u)G^>_{(0),jj} (u,w)   f^{\bar t}_{ji} (w) G^<_{(0),ii} (w,u)\Big).
\label{source2}
\end{align}
This agrees with \cref{source0}.

%%%%%%%%%%%%%%%%%%%%%%%%%%%%%%%%%%%%%%%%%%%%%%%%%%%%%%
 
\subsection{CP even and odd sources}
\label{s:valid}

The integral in the source \cref{source2} cannot be evaluated analytically for a generic background $f(x)$. To proceed it is assumed that the background is slowly varying, which in addition allows to identify to dominant contribution to the CP even and CP odd parts of the source.  This expansion is not the usual derivative expansion, in the sense of an expansion in powers of the diamond operator in the Wigner space KB equations \cref{kinetic}; indeed \cref{source2} {\it
  is} the leading order term, and -- as follows from \cref{der_vanish}
-- the only contribution from the diamond expansion. What is done
instead is that the mass insertion $f(w)$ in the source
\cref{source2}, with $w$ the coordinate that is integrated over, is
expanded as\footnote{At next-to-leading order, i.e. for the term with one derivative, only the time component $\lambda =0$ yields a non-zero result.}
\be
f(w) = f(u) + \partial_\lambda f(u) (w-u)^\lambda + {\cal O}(\partial^2).
\label{via_exp}
\ee
The $u$-dependent functions can then be taken out of the integral, and the resulting integral becomes manageable. The leading and next-to-leading terms in this expansion give a CP even and CP odd contribution to the source. For the left flavor
 \begin{align}
 S_L^{\textsc{CP}}
   &= 2 |\delta m^2|^2\int \dd^4 w \, {\rm Re} \Big(G^<_{(0),RR} (u,w)  G^>_{(0),LL} (w,u) -G^>_{(0),RR} (u,w)   G^<_{(0),LL} (w,u)\Big)  \\
   S_L^{\textsc{CPV}}&= 2 {\rm Im}\[\delta m^2 (\delta m^2)^{*'}\]  \times \nn \\
   & \hspace{1cm}\int \dd^4 w \, {\rm Im}\[ (w-u)\(G^<_{(0),RR} (u,w)  G^>_{(0),LL} (w,u) -G^>_{(0),RR} (u,w)   G^<_{(0),LL} (w,u)\) \] ,\nn 
 \end{align}
where we used \cref{Pi_0} for $f(u)$. The source for the right flavor is $S_R = -S_L$.
 
The expansion \cref{via_exp} is only valid if the background varies on
much larger length scales then the length scales dominating the
integral in the expressions above, otherwise the full spacetime
dependence of the mass should be taken into account. The mass
insertion is a function of the background bounce solution. For 
large bubbles curvature effects can be neglected, and the bubble is
generically well approximated by a kink solution, which in the bubble wall
rest frame takes the form \cite{John:1998ip}
\be
v_b (u) \approx \frac{v_N}{2} \(1 + \tanh( \frac{u^0-u^3}{L_w})\).
\ee
Here $v_N$ is the vev in the bubble interior, $L_w$ the bubble wall width,
$u^0= v_w t$ with $v_w$ the bubble wall speed, and $u^3$ the
distance from the center of the wall.

The background thus varies on length scales of the bubble wall width.  The typical scales dominating  the integral can be estimated from the explicit expressions for the source terms; here we look at the CP-odd source which has been calculated to give \cite{Riotto:1995hh,Riotto:1997vy,Lee:2004we}
\begin{align}
S_L^{\rm CPV} 
&\propto \int \frac{k^2 \dd k }{\omega_L
  \omega_R} \Im\bigg[ 
 \frac{\(n_B(\E_L^*) - n_B(\E_R) \)}{ (\E_L^*-\E_R)^2}
+\frac{\(n_B(\E_L) + n_B(\E_R) \)}{(\E_L +\E_R)^2}
                \bigg] \nn \\
  &\propto \int \frac{k^2 \dd k }{\omega^2} \Re \[\frac{\e^{\omega/T}}{\Gamma_T (1+\e^{\omega/T})^2} + {\cal O}(\Gamma_T)\]_{\bar m_{T,L}^2=\bar m_{T,R}^2},
\label{Sfinal}    
\end{align}
with $n_B(\E) $ is the Bose-Einstein distribution, and $k = |\vec k|$
the modulus of the three-momenta. The complex energies are defined via
$\E_j =\omega_j - i \Gamma_j$ with
$\omega_j=\sqrt{k^2+\bar m_{T,j}^2} $ with
$\bar m_{T,j}^2 = m_{j}^2+m_{T,j}^2$ the sum of the diagonal
zero-temperature mass and the thermal mass, and $\Gamma_{T,j}$ the
thermal decay width. On the 2nd line in \cref{Sfinal} we took the
limit of degenerate masses $\bar m_{T,L}^2 =\bar m_{T,R}^2 \equiv \bar
m_T^2$.  This
expression shows that the integral over three momentum is cutoff by
the mass term in $\omega$ and the integral varies over $|\vec k| \sim \bar m_T$,
corresponding to a length scale $\bar m_T^{-1}$.

To find the typical $k^0$ momenta, and the typical scale of the conjugate $u^0$ variation, it is useful to see how the denominators in the first line of \cref{Sfinal} arise, namely from integrals of the form
\be
\int_0^\infty \dd u^0 \, u^0\e^{i k^0 u^0} = -\frac{1}{k_0^2}, \qquad
k_0 = \E_L^*-\E_R, \,\E_L+\E_R,
\ee
with the specific $k_0$ values determined from contour integrals and the poles of the propagators. The $u^0$-integral is cutoff by the thermal width $\Im(k_0) \propto \Gamma_T$.

We thus conclude that the background changes slowly compared to the
length scales dominating the intergral, and can be safely approximated
as nearly constant,
if\footnote{Refs. \cite{Riotto:1995hh,Riotto:1997vy} cite the milder
  bound $L_w \gg v_w \Gamma^{-1}_T$.}
\be
L_w \gg \max\[ \Gamma^{-1}_T, \bar m_T^{-1}\],
\label{valid}
\ee
where the two terms on the right hand side come from the constraints on the temporal and spatial variation respectively.  If \cref{valid} is violated the expansion \cref{via_exp} breaks down, and the resulting expressions for the source \cref{Sfinal}  cannot be trusted.

The source term \cref{Sfinal} diverges for degenerate flavor-diagonal masses in the limit $\Gamma_T \to 0$. This has been noted before, and questions on the physical origin of this divergence were raised \cite{Cline:2020jre}. We see now that the expansion \cref{via_exp} used to derive this result breaks down when the inverse thermal width becomes larger than the bubble wall width $\Gamma_T^{-1} > L_w$. This cuts off the divergence for any finite bubble size.

Although for simplicity we have concentrated on the scalar model, the derivation in VIA of the source for fermionic systems is very similar, and the resulting source terms have a very similar structure to \cref{Sfinal}.  In particular, the same expansion \cref{via_exp} is made, and the same bound on the validity \cref{valid} applies. We can thus apply this limit on the use of the VIA source also for fermionic systems. As an explicit example, consider that the CP violation originates from modified Yukawa interactions
\be
{\cal L}_{\rm int} =- \delta m(z)\bar\psi_L \psi_R - \delta m(z)^* \bar \psi_R \psi_L,
\ee
with $\psi_i$ the left- and right-hand chiralities of a Standard Model
(SM) fermion.\footnote{The left and right chiralities are the
  equivalent of the left and right flavors of the scalar model, with
  the mass insertion off-diagonal in left-right `flavor' space. } In the Standard Model
$\delta m$ is just the mass term arising from the Yukawa coupling to
the Higgs, but to obtain sufficient CP violation requires new physics
corrections to this, which we leave unspecified.  Including thermal
corrections, crucial for the VIA source term,  the fermion propagator
is dressed with a flavor diagonal thermal mass and thermal width
$m^2_{T,i}$ and $\Gamma_{T,i}$ with $i=L,R$. The thermal corrections
are of the order $m^2_T \sim \alpha T^2$ and $\Gamma_T \sim \alpha T$,
with $\alpha \sim \alpha_3$ the QCD coupling for quarks and $\alpha
=\alpha_2$ the electroweak coupling for leptons. The VIA results are
then valid, see \cref{valid}, for
\be
L_w T \gg \alpha^{-1} \sim \left\{ \begin{array}{rl} 10, & \qquad
                                                            {\rm for} \;{\rm quarks} ,\\
                                                       30, &\qquad  {\rm for} \;{\rm leptons},
                               \end{array} \right.
\ee                            
which requirers thick bubble walls.

%%%%%%%%%%%%%%%%%%%%%%%%%%%%%%%%%%%%%%%%%%%%%%%%%%%%%%%
%%%%%%%%%%%%%%%%%%%%%%%%%%%%%%%%%%%%%%%%%%%%%%%%%%%%%%%
%%%%%%%%%%%%%%%%%%%%%%%%%%%%%%%%%%%%%%%%%%%%%%%%%%%%%%%
\section{Discussion}

In this paper we have calculated the source term at second order in
the vev insertion approximation (VIA) from the self-energy and Green's
functions both calculated at first order. Our approach agrees with the
known results, but has the advantage that it shows clearly the
equivalence of expanding the transport equations in terms of mass
perturbations or in self-energies.  The source term then derives, as
expected from this equivalence, from the same term in the kinetic
Kadanoff-Baym equation \cref{WKB} as semiclassical source terms based
on a gradient expansion. In particular, both the semiclassical WKB
force and flavor mixing forces can be derived from this term, see
\cite{Garbrecht:2018mrp} and references therein.

Despite their common origin the VIA and semiclassical source terms
have different parameteric dependence, and it thus appears that they
are distinct sources \cite{Garbrecht:2018mrp}. The VIA source
crucially depends on the CP-even phase provided by the thermal width
in the dressed propagator \cref{spectral}, and in fact vanishes in the
limit $\Gamma_T \to 0$ for non-degenerate flavor diagonal masses
$m_L^2 \neq m_R^2$, while the semiclassical source terms survive this
limit. It is clear that resumming the mass insertions in VIA cannot
yield the semiclassical force, and vice versa, the semi-classical
force can not be derived from an expansion in mass perturbations of
the VIA source. Their difference derives thus not so much from the
expansion in mass perturbations, but rather can be traced back to
taking different moments of the Kadanoff-Baym equation and a different
derivative expansion.

In the semiclassical scheme the Wigner space KB-equations are
integrated over $k^0$-momenta, to obtain Boltzmann equations for the
phase space densities. The source arises at first order in the
gradient expansion, i.e., in the expansion in diamond
operators.\footnote{For bosons, the WKB source at first order in the
  gradient expansion is CP conserving, and a CP violating source can
  only originate at higher order \cite{Konstandin_2005,Cline:2000nw}.} Instead, in the
VIA scheme the KB-equations are integrated over four-momenta,
which projects out all higher order terms in the the diamond expansion
\cref{der_vanish}, and thus also gets rid of the semiclassical source. The
resulting equations are formulated in terms of number currents (and
using the diffusion Ansatz, number densities) rather
than phase space densities.

To evaluate the integral in the VIA source term \cref{source2} another
derivative-type expansion is performed \cref{via_exp}, which assumes
the background varies slowly compared to the length scales dominating
the source integral.  We estimate that this expansion is valid, and
the VIA results can be trusted, for bubble widths
$L_w \gg \max\[ \Gamma^{-1}_T, \bar m_T^{-1}\]$, with $\bar m^2_T,$
the flavor diagonal mass including thermal corrections, and $\Gamma_T$ the
thermal width. The thermal corrections arise from the plasma
interactions.

The VIA source has a resonance for equal mass terms for the two
flavors in the free Lagrangian, as can be seen from the explicit
expression \cref{Sfinal}. This suggests it arises from (resonant)
flavor oscillation dynamics. We would then expect the resonance to
shift by order $\delta m^2$-corrections if the mass insertions are
resummed and the difference between flavor and mass eigenstates fully
accounted for.  As the VIA expansion only converges for small enough
mass perturbations, these corrections are relatively small. The VIA
source was calculated with four mass insertions in
\cite{Postma:2019scv}; for degenerate flavor diagonal masses this
next-to-leading order term is small if
\be
{\rm bosons}:\;\; \delta m^4 \ll \bar m_T^2 \Gamma_T^2,
\qquad
{\rm fermions}:\;\;  \delta m^2\ll \frac{\bar m_T^2  \Gamma_T^2}{
  (\bar m_T^2+\Gamma_T^2)}
\label{dm_small}
\ee
for bosons and fermions respectively
\cite{Postma:2019scv}.\footnote{For top quarks the vev insertion
  expansion breaks down as \cref{dm_small} is not satisfied, and the
  vev insertions must be resummed.}  The resonance is cut off by the
thermal width.

It will be interesting to further explore the relation between the VIA
source and semiclassical (resonant) flavor sources derived in an
approach based on a gradient expansion
\cite{Cirigliano:2011di,Cirigliano:2009yt}. Although the semiclassical
approach can be derived withouth thermal corrections, the divergence
in the degenerate mass limit is unshielded (see e.g. eq. (174) in
\cite{Garbrecht:2018mrp}), and thermal corrections should be included
for theoretical consistency.  To further assess and improve the
estimate on the validity of the VIA results, it will also be useful to
explicitly calculate the next term in the derivative expansion
\cref{via_exp}. In addition, one could try to devise alternative
methods to evaluate the source term for small bubble wall widths, to
bypass the expansion completely. This is left for future work.

\section*{Acknowledgments}
The author thanks Jorinde van de Vis and Jordy de Vries for very useful discussions. This work was funded by the Netherlands Organization for Scientific Research (NWO).

%%%%%%%%%%%%%%%%%%%%%%%%%%%%%%%%%%%%%%%%%%%%%%%%%%%%%%%
%%%%%%%%%%%%%%%%%%%%%%%%%%%%%%%%%%%%%%%%%%%%%%%%%%%%%%%
%%%%%%%%%%%%%%%%%%%%%%%%%%%%%%%%%%%%%%%%%%%%%%%%%%%%%%%

\appendix
\section{Diamond operator}
\label{A:diamond}

To rewrite the KB equations in Wigner space we use the following identities for the diamond operator
\begin{align}
\int \dd^4(u-v) \e^{i k.(u-v)} \int \dd^4 w A(u,w) B(w,v) &= \e^{-i \diamond}  \(A(k,x) B(k,x) \)
    \nn \\
  \int \dd^4(u-v) \e^{i k.(u-v)} A(u) B(u,v) &= \e^{-i \diamond}  \(A(x)B(k,x) \)
 \nn \\
                                               \int \dd^4(u-v) \e^{ik(u-v)} \(\partial^A_u+2\partial^B_u\) A(u) B(u,v)
                                                          &= \e^{ -i\diamond}  \(\partial^B_x -2ik\) A(x)  B(k,x)
\nn \\
 \int \dd^4(u-v) \e^{ik(u-v)} \(\partial^A_v+2 \partial^B_v\)  B(u,v) A(v)
&= \e^{ -i\diamond}  \(\partial^B_x +2ik\) A(x)  B(k,x)                                                            
                                               \label{diamond}
\end{align}
To derive the first identity above, we rewrite the left hand side in terms of the collective coordinate $x = \frac12(u+v)$ and relative coordinate $r= u-v =r_1+r_2$ with $r_1 =u-w$, and $r_2 =w-v$.  
\begin{align}
A(u,w) &=\bar  A(r_1, x+\frac12 r_2) = \e^{\frac12 r_2 \partial_x^A } \bar A(r_1, x) \nn\\
B(w,v) &=\bar  B(r_2,x -\frac12 r_1) = \e^{-\frac12 r_1 \partial_x^B } \bar B(r_2, x)
\end{align}
where in the last step we used that $\partial_x$ is the generator of translation, and the superscript denotes the term the derivative is acting on. 
Then
\begin{align}
 & \int \dd^4(u-v) \e^{ik(u-v)} \int \dd^4 w \,A(u,w) B(w,v) \nn \\
& \hspace{2cm}
=\int \dd^4 r_1\, \e^{(ik-\frac12\partial_x^B )r_1 } \bar A(r_1,x)
                                                                         \int \dd^4 r_2 \,\e^{(ik +\frac12 \partial_x^A) r_2 }\bar B(r_2,x) \nn \\
&  \hspace{2cm}= A(k+\frac{i}{2} \partial_x^B, x) B(k-\frac{i}{2} \partial_x^A, x)
  =\e^{-i \frac12\(\partial_x^A \partial_k^B  -\partial_x^B \partial_k^A\) }A(k, x) 
    B(k, x) \nn \\
  & \hspace{2cm} = \e^{-i \diamond} A(k, x) 
B(k, x).
\end{align}
where in the penultimate step we used that $\partial_k$ is the generator of translations in momentum space. The derivation of the other identities in \cref{diamond} can be done in a similar way.

The derivative expansion $\e^{-i\diamond} =1- i\diamond+...$. transformed back to position space becomes
\begin{align}
  \int \frac{\dd^4 k}{(2 \pi)^4}  \e^{-i\diamond}  A(k,x) B(k,x)
              &=
      \int \frac{\dd^4 k}{(2 \pi)^4}   \int \dd^4 r \int \dd^4r' \e^{-\frac{i}{2}\(\partial_x^A  \cdot\partial_k^B
-\partial_k^A \cdot \partial_x^ B\) }\e^{ik(r+r')} \bar A(r,x) \bar  B(r',x)
 \nn\\
   &=
\int \dd^4 r \,
     \e^{ r \cdot\(\partial_x^A  + \partial_x^ B\) }\bar A(r,x) \bar B(-r,x)
     \nn\\
 % &=
% \int \dd^4 v\,
%      \e^{(u-w) \cdot\(\partial_u  + \partial_w\) } A(u,w)  B(w,u)
%    \nn\\
  &=
\int \dd^4 w
    \[1+ (u-w) \cdot\(\partial_u  + \partial_w\) +{\cal O}( \partial^2) \]A(u,w) B(w,u).
    \label{invWig}
\end{align}
Note that the leading order derivative expansion corresponds to the limit $v \to u$ in position space
\begin{align}
  \int \dd^4 w
   A(u,w) B(w,v) \Big|_{\textsc{LO}}&=
                  \int \frac{\dd^4 k}{(2 \pi)^4}  \e^{-i\diamond}  A(k,x) B(k,x) \Big|_{\textsc{LO}}
                  =    
\int \dd^4 A(u,w) B(w,u).
    \label{lim_uv}
\end{align}

\bibliographystyle{jhep} 
\bibliography{myrefs}

\end{document}